\documentclass[reqno, final, longbibliography,notitlepage]{article}
  \usepackage[noblocks]{authblk}

\usepackage[a4paper, margin=2.5cm]{geometry}
\usepackage[usenames,dvipsnames]{color}
\usepackage{graphicx}
\usepackage[utf8]{inputenc}

\usepackage{bbm}

\newcommand\R{\mathbbm{R}}

\newcommand\KS{\mathcal{KS}}

\usepackage{hyperref}
\hypersetup{ colorlinks=true,citecolor=cyan, urlcolor=blue,breaklinks}

\usepackage{amsmath,amsfonts,amssymb,amsthm}
\newtheorem{theorem}{Theorem}%[section] 
\newtheorem{example}{Example}%[section]
\begin{document}
%\begin{frontmatter}
\title{ Tangent bundle geometry from dynamics: application to the Kepler problem}

\author[1,2]{J. F. Cari\~nena  }

\affil[1]{Departamento de F{\'{\i}}sica Te\'orica, Universidad de Zaragoza, 
  Pedro Cerbuna 12, E-50009 Zaragoza, Spain}
\affil[2]{IUMA, Campus San Francisco, Universidad de Zaragoza,  50009 Zaragoza, SPAIN}

\author[1,3]{J. Clemente-Gallardo}
\affil[3]{Instituto de Biocomputaci\'on y F{\'{\i}}sica de Sistemas 
 Complejos (BIFI), Universidad de Zaragoza, Mariano Esquillor s/n, Edificio 
 I+D, E-50018 Zaragoza, Spain}

\author[1,3]{J. A. Jover-Galtier}

\author[4,5]{G. Marmo}
\affil[4]{ Dipartimento di Fisica, Universit\`a Federico II di Napoli
   80126-Napoli (ITALY)}
\affil[5]{INFN Sezione di Napoli, 80126-Napoli (ITALY)}

\maketitle

\abstract{
	In this paper we consider  a manifold with a dynamical vector
        field and inquire about the possible tangent bundle structures
        which would turn the starting vector field into a second order
        one. 
The analysis is restricted to manifolds which are diffeomorphic with
affine spaces. In particular, we consider the problem in connection
with conformal vector fields of second order  and apply the
procedure to vector fields conformally related with the harmonic
oscillator (f-oscillators) . We select one which covers the vector
field describing the Kepler problem. 
}

%\keywords{complete vector fields;
%second order vector fields; tangent bundle structures; f-oscillators; Kepler dynamics}

%\end{frontmatter}
\section{Introduction}

Often, for various reasons (reduction, quantum-to-classical
transition, repar\-ame\-trization, statistical models, etc) we  are
dealing with dynamical systems on carrier spaces where a clear
identification of ``positions" and ``velocities" (or ``momenta") need not be
available.  It is therefore meaningful to investigate if and when it is possible to 
identify ``positions" and ``velocities" or  ``momenta"
Equivalently, if we are given a dynamical system
 described by a vector field $\Gamma$ on some carrier manifold $M$, we
 would like to investigate whether and when the carrier manifold may be
 given a tangent bundle structure which turns  the given vector field
 into a second order vector field. This could also be a way, for instance,  to investigate
 the inverse problem of the calculus of variations in a greater
 generality. 
 
In order to avoid topological obstructions, we shall consider in this
paper only the simplest situation: $M=\mathbb{R}^{2n}$. If we
are aiming, for instance, at the Lagrangian case,  our goal would be  to identify
a submanifold $\mathbb{R}^{n}\subset M$ which would allow to write
$M\approx  T \mathbb{R}^{n}$, in such a way that $\Gamma$ becomes a second
order differential equation vector field with respect to the tangent bundle
structure.  

The simplest   example is provided by the free particle  in $\R^3$,
even though it is described by the free Lagrangian  
$$
{L}=\frac{1}{2}\langle \dot{\vec{r}},\dot{\vec{r}}\rangle, 
$$
the reduction of the dynamics along the radius is not described by the
``reduced Lagrangian" (see \cite{Carinena2007c}) but a simple minded
reduction would have the wrong sign in the ``centrifugal  
potential". Therefore the Lagrangian description of the reduced
dynamics needs to be found by means of the inverse problem. In this
way we would find the correct ``minus sign" in the centrifugal
potential. 

In other reductions the reduced carrier space may turn out to have odd dimension and therefore the identification of ``positions" and ``velocities" would not be possible.
 Analogous situations may arise within the quantum-to classical transition where the classical limit need not be directly a phase space. This would occur, for instance for particles with spin.

A similar situation was studied from the Hamiltonian perspective in
\cite{Marmo1976} under the name of ``$Q$-dynamical systems". Such objects
were introduced as the flow under the diffeomorphism group of  a
Hamiltonian vector field $X$ with respect  to a given symplectic form
$\omega$.  This starting symplectic form was considered to be the
canonical symplectic form of the cotangent bundle $T^*Q$ for a given
choice of the configuration space $Q$.  However, by performing a
generic transformation by means of a non-canonical diffeomorphism 
 which would be a symmetry for the dynamics, as the symplectic
 structure would not be preserved, we would loose the  separation
 between ``positions" and conjugate momenta. 

In conclusion we would face the problem of reconstructing a tangent or
a cotangent bundle structure after a transformation by a generic  
diffeomorphism which would not be  a tangent bundle automorphism.

It should be noticed that this situation would already occur from one
Galileian frame to a different one, expressing the fact that ``zero
velocity" or ``zero motion" of the  
tangent bundle cannot have an absolute meaning.  Indeed, let us
consider for simplicity the vector field $\Gamma$ of a free motion on
$\R^{2n}$. Thus, we  assume that there exists coordinate functions $\{
\eta^{j}, \xi^{k}\}$  satisfying 
\begin{equation}
  \label{eq:15}
  \mathcal{L}_{\Gamma}\eta^{j}=\xi^{j}; \qquad
  \mathcal{L}_{\Gamma}\xi^{j}=0 \quad \forall j.
\end{equation}
Clearly, if we choose as coordinates
\begin{equation}
  \label{eq:16}
  \eta'^{k}=\eta^{k}+f^{k}(\vec \xi), \qquad \xi'^{j}=\xi^{j}
\end{equation}
for $f^{k}$ arbitrary smooth functions, the coordinates $(\eta', \xi')$ 
share the same properties as the original ones with respect to
$\Gamma$.  Hence, we see that there is some freedom in the choice of
coordinates of the submanifold representing the ``positions''.  If the functions
$\xi^{j}$ are such that the set $\vec \xi=0$ defines a smooth submanifold
of $\R^{2n}$, say $Q$, then our original manifold is diffeomorphic
with $TQ$.  But Galileian relativity would  change the zero section of
that bundle, since velocities would be changed by a constant vector.
We conclude thus that for a free motion to be modeled as a SODE and
to take into  account Galilean relativity, we must consider different
tangent bundle structures, one particular structure being 
associated with every  chosen frame.

 As it is well-known a tensorial description of a tangent bundle
 structure is provided by two tensor fields, $S$, soldering (1,1)
 tensor field, and a Liouville vector field, also called ``partial
 linear structure". 
More specifically, it is known
\cite{josegeometry,crampin1983defining,Crampin1983,Crampin1985,Filippo1989,Grifone1972a,Grifone1972b}
that  the geometry of a tangent bundle is encoded in two tensors:
\begin{itemize}
	\item the vertical endomorphism $S$, a $(1,1)$ tensor defined on the
	tangent bundle $TM$ on a manifold $M$
	which in the natural coordinates $(q^{k}, v^{k })$ reads
	\begin{equation}
	\label{eq:94}
	S=dq^{k}\otimes \frac{\partial}{\partial v^{k}},
	\end{equation}
	\item and the Liouville vector field, which encodes the dilations along
	the fibers and hence their linear structures.
	The corresponding local expression reads
	\begin{equation}
	\label{eq:95}
	\Delta=v^{k}\frac{\partial }{\partial v^{k}}.
	\end{equation}
\end{itemize}

By construction, $S^{2}=0$ and ${\mathcal L}_{\Delta}S=-S$. Both tensors
allow us to define Second Order Differential Equations (SODE) to be
those vector fields $X\in \mathfrak{X}(TM)$ which satisfy
\begin{equation}
\label{eq:96}
S(X)=\Delta.
\end{equation}
This relation shows very clearly that the condition for a vector field
to be a SODE depends on both $S$ and $\Delta$.

A translation along each fiber of the tangent bundle would not alter
the soldering tensor but would change the dilation vector field. The
vector field  $\Delta= (v^i+c^i(q))\, \partial/\partial v^i$ also
satisfies 
$\mathcal{L}_\Delta S=-S$, but then the condition $S(\Gamma)=\Delta$
would not hold true for vector fields which were second order with
respect to the previous structure.
%be the condition for second order differential equation
%vector field. 

Furthermore, given a regular Lagrangian function $L$,
 Euler-Lagrange equations  
which are implicit differential equations, may be solved in terms of
the solutions of a second order vector field $\Gamma$, if it exists,
that satisfies the following equation 
\begin{equation}
\label{eq:97}
{\mathcal L}_{\Gamma}\theta_{L}-dL=0,
\end{equation}
where $\theta_{L}$ is the 1-form defined by the dual map of the vertical
endomorphism (denoted by $S^{*}$) as
\begin{equation}
\label{eq:98}
\theta_{L}=S^{*}(dL)=dL\circ S.
\end{equation}
 
 The   main result can be summarized as follows (see \cite{josegeometry, Filippo1989}):
 
\begin{theorem}
	\label{theoremTM}
	 A  differential manifold $M$  may be endowed with a
         tangent bundle structure if and only if there exist a
         complete vector field $\Delta$, whose zeros define a
         submanifold $Q$ of $M$,  and a $(1,1)$ tensor field $S$
         satisfying the following conditions: 
\begin{itemize}
	\item $\mathrm{Ker}S=\mathrm{Im} S$, i.e. $S^{2}=0$
	\item $\Delta\in \mathrm{Im}S$, what implies that $S(\Delta)=0$
	\item ${\mathcal L}_{\Delta}S=-S$
	\item $N_{S}=0$, where $N_{S}$ stands for the Nijenhuis torsion for $S$
	\item the limit of the flow of $\Delta$, $\lim_{t\to
            -\infty}\Phi_t^\Delta(p)$  exists for any $p\in M$. 
\end{itemize}
$S$ and $\Delta$  uniquely determine a tangent bundle structure for $M$. Indeed,  it is
possible to  identify the submanifold $Q$ as the base manifold which makes $M$
diffeomorphic to $TQ$.   When considering 
adapted coordinates $(q^{i}, v^{j})$ on this $TQ$, the local expressions of the
tensors become (\ref{eq:94}) and (\ref{eq:95}).
	 
\end{theorem}

Different choices of these tensors define different tangent bundle
structures for the manifold $M$.  In this paper  we are going to
study the general conditions to determine a suitable tangent bundle
structure that makes a given vector field $\Gamma$ a SODE on
$M=\R^{2n}$.

We will
also focus our attention  on two particular situations where such a construction 
produces interesting properties: the definition of a new vector field
conformal to the given $\Gamma$ (as for instance when we try to
regularize it to make it complete)
and the deformation of a vector field by a function, in particular in
the form of an $f$--oscillator introduced in \cite{Man'ko1997}.  As an
important  application, we will  consider the regularization of the Kepler
problem, and we will see that the redefinition of the tangent bundle
structures will allow us to define a suitable diffeomorphism of its
space of motions and the space of motions of a $f$--oscillator.  Such
mapping circumvents the obstruction arising from the energy-period
theorem (see \cite{gordon1969relation,Gordon1975, Lewis1955}) to
identify both systems at the phase-space level.

The scheme of the paper is as follows. Section
\ref{sec:conf-relat-vect} introduces the case of the
repara\-metrization of a vector field and the type of conformal
factor required to make the reparametrized vector field complete.
Finally, we will discuss how this process 
affects the set of periodic orbits.  In Section
\ref{sec:$f$--oscillators} we will consider the deformation of vector
fields by introducing the notion of $f$--oscillator. In particular, we
introduce for the first time here the Lagrangian version of that
construction and show how the resulting system takes also the form of
a conformal vector field with the conformal factor being a function of
the oscillator energy. Then, Section \ref{sec:altern-tang-bundle} presents
the main result of the paper: given a  vector field
$\Gamma'$ on some carrier space $M$,  is it possible to define a
suitable tangent bundle structure on $M$ which makes $\Gamma'$ a SODE?
We will obtain the generic properties for that to happen and present
some relevant examples. In particular we study the case where
$\Gamma'$ is a conformal vector field with respect to a SODE vector
field in  some original tangent bundle structure and look
for the conditions for the existence of a new structure with respect
to which the new vector field $\Gamma'$ is a SODE. We will apply this
to the regularized Kepler problem and the $f$--oscillator problem and
we will prove that the new SODE vector fields are formally identical
and equal to a harmonic oscillator with a frequency which depends on
the point.  Finally,
in Section \ref{sec:space} we will prove that the similarities of the
resulting SODE vector fields of two systems allow us to define an
invertible mapping   of the corresponding spaces of motion  and apply that
to the case of the Kepler problem and the $f$--oscillator. Thus we
will prove that it is possible to identify a suitable deformation
function $f$ to deform a harmonic oscillator in such a way that the
resulting SODE vector field has the same frequencies as the SODE of
the regularized Kepler problem, allowing us to define a one-to-one
relation of their respective motion spaces. Finally, Section
\ref{sec:conclusions}  summarizes the 
main results of the paper and discuss some of its possible extensions.

\section{Conformally related vector fields: 
reparametrization and  completeness} 
\label{sec:conf-relat-vect}

The need for reparametrization of a vector field may come from
different situations. One possible case is the need of defining a
complete vector field associated to a given dynamics, as it happens,
for instance, in the case of the closed orbits of the Kepler problem.
Another situation requiring regularized vector fields is the case of
quantization, where systems whose classical dynamical vector field is
not complete may give rise to quantum Hamiltonian operators which are
not self-adjoint and therefore nonphysical (see \cite{Zhu1993a}). In
this case the appropriate  identification of the corresponding conjugate
variables becomes very important.

  Hence in this Section we are going to discuss a few
properties of the regularization mechanism, and, in particular, the
implications at the level of the period of the  orbits of the
vector fields, which is our main concern.

\subsection{Definition and effects on tensors}
\label{sec:repar-vect-fields}
The notion of conformally related  differential operators is of great importance in many
areas of Physics. We will summarize now the most relevant ones for
vector fields, and we address the interested reader to  the work by
Palais \cite{Palais1957} or to \cite{josegeometry} for more details.

We say that two vector fields $X$ and $ \Gamma$  on a differential manifold $M$ 
are conformally related  vector
fields if there exists a nowhere vanishing differential function
$f\in C^{\infty}(M)$ such that $\Gamma=fX$. This defines an equivalence relation on the set
of vector   fields of a differential manifold.
The dynamics associated to $\Gamma$ is very
similar to the dynamics associated to $X$, but it also has some
interesting differences. It is straightforward to prove, for instance,
that  $X$ and $\Gamma$ admit the same algebra of constants of the motion. Indeed, a function 
$g\in C^{\infty}(M)$ is a  constant of motion for $X$, i.e $ {\mathcal L}_{X}g=0$, if and only if  $\mathcal {L}_{\Gamma}g= 0$, because
${\mathcal L}_{\Gamma}g={\mathcal L}_{fX}g=f{\mathcal L}_{X}g$.

However, the algebra of invariant tensor fields under $X$ does not
coincide with the algebra of invariant tensor fields under $\Gamma$.
Indeed,
if we consider for instance a 1--form $\alpha\in \bigwedge^{1}(M)$
which is preserved by the 
vector field $X$, it may happen that $\Gamma$ does not preserve
it, because 
\begin{equation}
\label{eq:45}
{\mathcal L}_{\Gamma}\alpha=(di_{\Gamma}+i_{\Gamma}d)\alpha=d(fi_{X}\alpha)+f(i_{X}d\alpha)= (i_{X}\alpha)\, df+ f{\mathcal L}_{X}\alpha,
\end{equation}
and  the term $ (i_{X}\alpha)\, df$ may be different from zero even
if ${\mathcal L}_{X}\alpha=0$. 
Analogously, if we consider a 2-form as $\alpha\wedge \beta$ for $\alpha, \beta\in
\Lambda^{1}(M)$, we obtain
\begin{equation}
\label{eq:49}
\mathcal{L}_{\Gamma}(\alpha\wedge
\beta)=(\mathcal{L}_{\Gamma}\alpha)\wedge \beta+\alpha \wedge
(\mathcal{L}_{\Gamma}\beta)=(i_{X}\alpha)\, df\wedge \beta
+(i_{X}\beta)\,\alpha\wedge df +f(\mathcal{L}_{X}(\alpha\wedge \beta)),
\end{equation}
and then $\mathcal{L}_{X}(\alpha\wedge \beta)=0$ does not imply $\mathcal{L}_{\Gamma}(\alpha\wedge \beta)=0$.

If we consider the case of a vector field, we find similarly:
\begin{equation}
\label{eq:47}
{\mathcal L}_{\Gamma}Y={\mathcal L}_{fX}Y=[fX, Y]=-Y(f)X+f[X,Y] ,\quad
\forall Y\in \mathfrak{X}(M), 
\end{equation}
and then $\mathcal{L}_{X}Y=0$ does not imply $\mathcal{L}_{\Gamma}Y=0$.

In case of a bivector field $Y\wedge Z$, for any  $Y,
Z\in \mathfrak{X}(M)$, the situation is similar, because 
\begin{align}
\label{eq:48}
{\mathcal L}_{\Gamma}(Y\wedge Z)=({\mathcal L}_{\Gamma}Y)\wedge Z+Y\wedge
({\mathcal L}_{\Gamma}Z)=& -Y(f)X\wedge Z-Z(f)Y\wedge X+ \nonumber \\ &f({\mathcal
	L}_{X}Y\wedge Z +Y\wedge {\mathcal L}_{X}Z) \nonumber \\
=&\left (Y(f)Z-Z(f)Y \right )\wedge X+f({\mathcal L}_{X}(Y\wedge Z) )
\end{align}
which shows that ${\mathcal L}_{X}(Y\wedge Z)=0$ does not imply $\mathcal{L}_{\Gamma}(Y\wedge Z)=0$.  

The extension to more general tensors is straightforward. 
We notice that the conformal factor has consequences on the invariance
properties of tensors fields, however these changes are simple to compute.
This fact will be important in the
following, since we will need to redefine the geometrical structures
of $M$ to define new ones which are preserved by the vector field
$\Gamma$.  

It is also immediate that the orbits of vector fields $X$ and $\Gamma$
coincide.  Indeed  vector fields
$X$ and  $\Gamma$ are tangent to the same curves, the only difference
being the parametrization chosen.  Therefore we can understand the
conformal factor as a kind of reparametrization of the curves. Consider an integral
curve $x(t)$ of $X$  parametrized by $t\in I\subset  \mathbb{R}$, i.e.                                                                                                                                                                                                                                                                                                                                                                   
\begin{equation}
\label{eq:9a}
\frac {d x(t)}{dt}=X(x(t)),
\end{equation}
and let $s(t)$ be a solution of the differential equation
\begin{equation}
\label{eq:8}
\frac{ds}{dt}=f(x(t)),
\end{equation}
which defines a good reparametrization because $f$ is of a constant sign.
If we consider the                                                                                                                                                                                                                                                                                                                                                                                                                                                            condition for the curve  $x$ to be an integral
curve of the vector field $\Gamma$ we find:
\begin{equation}
\label{eq:9}
\frac {d x(t)}{dt}=\Gamma(x(t))=f(x(t))X(x(t)).
\end{equation}
It is immediate to see that by considering the reparametrization $s(t)$,
the flow satisfies:                                                                       
\begin{equation}
\label{eq:10}
\frac {d x(s)}{ds}=X(x(s)).
\end{equation}
Therefore the integral curves of $X$ and $\Gamma$ are the same
submanifolds of $M$, but with a different parametrizations.

\subsection{Completeness of a conformally-related vector field } 

In particular, we shall be interested in a conformally related  vector field
$\Gamma=fX$  which is complete, while $X$ is not. Such a
construction can be used in any paracompact manifold. Indeed, 
given a vector field $X$ on a paracompact manifold $M$, there always  exists
a strictly positive function $f\in {\mathcal F}(M)$, of the same
differentiability class as $X$ such that
$\Gamma=f\,X$
is a complete vector field.

To prove this assertion one notices that, due to the paracompactness
of $M$, there exists a proper function $g\in {\mathcal F}(M)$ of class
$C^{s}$ on $M$. Let us consider thus 
\begin{equation}
\label{eq:80}
f=\exp \left ( - \left ( {\mathcal L}_{X}g\right )^{2}\right ); \qquad \Gamma=fX
\end{equation}

It is immediate that
\begin{equation}
\label{eq:81}
| {\mathcal L}_{\Gamma}g|=\left | \exp \left ( -\left (
{\mathcal L}_{X}g \right )^{2} \right ) ({\mathcal L}_{X}g)\right |\leq 1
\end{equation}
on $M$. Therefore, if we denote by $\gamma$ an integral curve of the
vector field $\Gamma$ defined on a bounded interval $I$, then
\begin{equation}
\label{eq:82}
\frac d{dt} (g\circ \gamma)=({\mathcal L}_{\Gamma}g)\circ \gamma,
\end{equation}
and hence
\begin{equation}
\label{eq:83}
\left | \frac d{dt} (g\circ \gamma) \right | \leq 1,
\end{equation}
on $I$. Then the image of $g\circ \gamma$ is bounded and hence the
image of $\gamma$ is relatively compact.
%\subsection{Conformally related vector fields and the energy-period theorem}

%Notice thus that
%WHILE THE METRIC IS NOT PROJECTABLE, SOMEHOW THE KINETIC ENERGY IS

\subsection{Conformally related  vector fields, Hamiltonian structures and  periodic orbits}
As we said above,  the unparametrized orbits of two conformally related vector fields
$X$ and $\Gamma=fX$ 
coincide, but, in general, the vector field $\Gamma$ will not preserve
the same geometric structures as $X$. Therefore the reparametrization
defined by  function $f$ is not compatible 
with other properties, such as the fact of being Hamiltonian (unless
$f$ is a central element of  Poisson algebra, for
the corresponding symplectic or Poisson structure ) or being
a Second Order Differential Equation (if the manifold is a tangent
bundle).  
 We will see later that  in a particular case 
the condition of being a SODE can be recovered by defining a new
tangent bundle structure on the same manifold.

In  case of Hamiltonian structures, if $X=X_{H}$ is a Hamiltonian
vector field associated with a function 
$H\in C^{\infty}(M)$ on a symplectic manifold $(M, \omega)$, the vector field 
$\Gamma=fX_{H}$ is not  a Hamiltonian vector field, unless  
\begin{equation}
\label{eq:84}
i_{\Gamma}\omega=i_{fX_{H}}\omega=fi_{{X_{H}}}\omega=f\,dH.
\end{equation}
Instead, $\Gamma$ is just a \textit{conformally Hamiltonian vector field}.
The interest on conformally Hamiltonian vector fields comes from the
possibility of relating their integral curves with those 
of the Hamiltonian vector fields via a reparametrization of
the curves.
In general the 1-form $f\, dH$ is not  exact and  $\Gamma$
is not a Hamiltonian vector field. Only if 
$$
df\wedge dH=0,
$$
 the  vector field $\Gamma$  is Hamiltonian. But even in this case a transformation of
this type cannot make $\Gamma$ complete if $X_{H}$ is not, because
 a function of the Hamiltonian re-scales the parameter by
a different constant value along each curve.

Notice that these properties are meaningful from the physical point of
view, since they are associated to quantities as the period of closed
orbits. 
%When $f$ is  a closed curve $\gamma$ and two vector
%fields $X$ and $\Gamma=fX$, having $\gamma$ as an  integral curve,\marginpar{impossible} 
%the
%periods of the two associated to each dynamical vector field are, in general, 
%different. 
From the reparametrization relation we introduced above, we can
consider the two vector fields $X$ and  $\Gamma=fX$ and the a curve
$\gamma$ satisfying
\begin{equation}
\label{eq:43}
\frac{d\gamma(t)}{dt}=\Gamma(\gamma(t))=f(\gamma(t)) X(\gamma(t)).
\end{equation}
If we consider the case where $f$ is a constant of the motion (as it
will  happen for $f$--oscillators), 
$f(\gamma(t))=K$ is constant for all $t$. Thus  we realize that the effect
on the curve $\gamma$ is just that the dynamics associated to $\Gamma$ runs
$K$ times faster (if $K>1$) than the dynamics associated to $X$. 
 It is straightforward to prove from Equation (\ref{eq:8}) that,
if the conformal factor function $f$ is constant along the curve, i.e.,
\begin{equation}
\label{eq:44}
f(\gamma(t))=K, \quad \forall t
\end{equation}
its period is scaled by $K^{-1}$.

Notice that, in general,  this re-scaling may differ from curve to curve, and
therefore we may have examples (e.g. the harmonic oscillator), which, once
transformed by a conformal factor, produce a set of circumferences as
integral curves but with periods which are different for each
curve. Therefore they cannot be considered as orbits of $SO(2)$ or $U(1)$ but should be considered orbits of $\mathbb{R}$.

This result is relevant if we aim to relate two Hamiltonian systems
with periodic orbits. It implies, for instance, that we cannot map
smoothly the set of solutions of a system with constant
period (as the harmonic oscillator) on the solutions of a system with
non-constant period (as for instance the Kepler problem).  It is
possible though to relate both models via a family of mappings
parametrized by the energy. Then, an orbit corresponding to a
certain energy in the Kepler problem is covered by  closed orbits
of the harmonic oscillator, but the mapping is not global: another
Kepler-orbit with a different energy is covered by 
oscillator-orbits by a different mapping  with
a different frequency (which depends on the energy). Still, from what we just
learned above,  we may also search for a
suitable conformal factor function $f$ which is a constant of the
motion for $X$, and that tunes the period of each orbit to make them
match exactly the function $T(E)$ which the energy-period theorem
predicts for the Kepler problem.  This is precisely what the
$f$--oscillator will do.

\section{Deforming dynamical vector fields: $f$--oscillators}
\label{sec:$f$--oscillators}

The notion of $f$--oscillator was introduced in \cite{Man'ko1997}  as a
procedure to define nonlinear coherent states by deforming the Hamiltonian
function of the isotropic harmonic oscillator.  This transformation changes the
energy-period relation  and allows us to
consider it as a candidate to relate systems with closed orbits which
have different energy-period relations  with  respect to that of the
harmonic oscillator. This property 
will prove to be essential in order to combine the results from
Section \ref{sec:kepler-problem} into a one-parameter family of orbits
of four-dimensional harmonic oscillators, 
each one with a different frequency, that are in 
correspondence with the set of closed orbits of the three-dimensional
Kepler problem.

%If we consider the non-isotropic HO everything keeps true.  The period
%of the original is still constant. When deformed that is still true if
%taken the total $H$. But if we consider sum of the squares $H_{k}^{2}$
%as deforming object, there are new orbits which are non-periodic. 

%The main reason is the
%reparametrization of the orbit which is associated to the conformal
%factor associated to the deformation, as we can see below.  Then, our goal in the
%rest of the paper will be to identify suitable conformal factors
%(i.e., suitable functions $f$) in order to relate the set of closed orbits of
%interesting models with the set of orbits of the $f$--oscillator. 

\subsection{The notion of $f$--oscillator}

Let us consider then an isotropic harmonic oscillator defined in
$n$--dimensions, i.e. the dynamics is defined on $\mathbb{R}^{2n}$,
parametrized by Darboux coordinates $\{(q^{k}, p_{k}\mid k=1, \cdots, n\}$  by the
Hamiltonian function
\begin{equation}
\label{eq:1}
H=\frac 12 \sum_{k=1}^{n}(p_{k}^{2}+(q^{k})^{2}), 
\end{equation}
with respect to the canonical symplectic form
\begin{equation}
\label{eq:2}
\omega=\sum_{k=1}^{n}dq^{k}\wedge dp_{k}.
\end{equation}
The corresponding Hamiltonian vector field reads:
\begin{equation}
\label{eq:3}
X_{H}=\sum_{k=1}^{n}\left ( p_{k}\frac{\partial}{\partial
	q^{k}}-q^{k}\frac{\partial}{\partial p_{k}}\right )
\end{equation}

The $f$--oscillator dynamics is defined by constructing a new Hamiltonian
function $\bar{H}$, which is obtained as the image of the original $H$ by a
function $f:\mathbb{R}\to \mathbb{R}$. Thus we consider
\begin{equation}
\label{eq:4}
\bar{H}=f(H)=f \left ( \frac 12 \sum_{k=1}^{n}(p_{k}^{2}+(q^{k})^{2}) \right )
\end{equation}

This Hamiltonian function produces a Hamiltonian vector field with
respect to the symplectic form (\ref{eq:2}):
\begin{equation}
\label{eq:5}
X_{\bar{H}}=\left .\frac{df(\xi)}{d\xi}\right |_{{\xi=H}}\sum_{k=1}^{n}\left ( p_{k}\frac{\partial}{\partial
	q^{k}}-q^{k}\frac{\partial}{\partial p_{k}}\right
)=\left .\frac{df(\xi)}{d\xi}\right |_{{\xi=H}}X_{H}.
\end{equation}
The resulting dynamics is no longer linear, although it still produces
periodic orbits. The main difference with respect to Equation (\ref{eq:3}) is that
the frequencies change from one to another 
level set of energy. Different orbits on different level sets will have
thus different periods, depending on the value of the derivative of
the function $f(H)$ when evaluated at the orbit. An important point
which will be relevant later is the fact that the level sets of the
$f$--energy (i.e. the set of points where $f(H)$ is constant) are
$2n-1$ spheres, as it  happens with the undeformed level sets. In
particular the group $SU(n)$ will be again a symmetry group of our
reparametrized Hamiltonian vector field.
The
deformation by the function $f$ affects only to the rate of change of
the radius from one to another  level set.

\subsection{Lagrangian formulation of a $f$--oscillator}
\label{sec:lagrangian-f}
The concept of $f$--oscillator was designed for and within the
Hamiltonian formalism. The original idea was to build a deformed
dynamics which was Hamiltonian with respect to an alternative
symplectic structure. 
We want to consider now an alternative point of view: we want to study
whether we can define a deformed description of a Lagrangian dynamics
in an analogous way. Thus consider the Hamiltonian of the harmonic
oscillator (\ref{eq:1})  and the corresponding Lagrangian description defined by
the Legendre transform: 
\begin{equation}
\label{eq:85}
{\mathcal F}H:T^{*}\R^{n}\to T\R^{n}
\end{equation}
where ${\mathcal F}H$ is defined as 
\begin{equation}
\label{eq:6}
{\mathcal F}H:(q^{k}, p_{k})\mapsto \left ( q^{k}, \frac{\partial
	H}{\partial p_{k}} \right ),
\end{equation}
which, in our case, is a global diffeomorphism.

The dynamical vector field
\begin{equation}
\label{eq:7}
X_{H}=\sum_{k} \left (  \frac{\partial H}{\partial
	p_{k}}\frac{\partial}{\partial q^{k}} - \frac{\partial H}{\partial
	q^{k}}\frac{\partial}{\partial p_{k}} \right )=\sum_{k} \left (
p_{k}\frac{\partial}{\partial q^{k}}-q^{k}\frac{\partial}{\partial p_{k}}\right )
\end{equation}
is mapped by ${\mathcal F}H_{*}$ onto the vector field $\Gamma\in
\mathfrak{X}(T\R^{n})$
\begin{equation}
\label{eq:86}
\Gamma=\sum_{k}\left ( v^{k}\frac{\partial}{\partial q^{k}}
-q^{k}\frac{\partial}{\partial v^{k}} \right ).
\end{equation}

$\Gamma$ is a SODE and it represents the dynamical vector field whose
integral curves are the solutions of the Euler-Lagrange equations for
the Lagrangian function 
\begin{equation}
\label{eq:87}
L=\frac 12 \sum_{k}\left ((v^{k})^{2}-(q^{k})^{2}\right ).
\end{equation}
This vector field can be given a symplectic formulation if we write
\begin{equation}
\label{eq:88}
i_{\Gamma}\omega_{H}=dE_{H},
\end{equation}
where the symplectic form $\omega_{H}$ is the pullback of the
canonical symplectic form $\omega$ on $T^{*}\R^{n}$ by the inverse of
${\mathcal F}H$, i.e.,
\begin{equation}
\label{eq:89}
\omega_{H}=(({\mathcal F}H)^{-1})^{*}\omega,
\end{equation}
and the energy function $E_{H}$ is the pullback of the Hamiltonian
function by ${\mathcal F}H^{-1}$:
\begin{equation}
\label{eq:90}
E_{H}=({\mathcal F}H^{-1})^{* }(H)=\frac 12 \sum_{k}\left ((v^{k})^{2}+(q^{k})^{2}\right ).
\end{equation}

If we consider the $f$--oscillator dynamics associated to the function
$H'$   defined $\bar H$  in Equation (\ref{eq:4}) and the vector field corresponding to
Equation (\ref{eq:5}), we know that the vector field $X_{H'}$ is  conformally related  to $X_{H}$,
$X_{H'}=f'X_{H}$. The image of the
vector field $X_{H'}$ under ${\mathcal F}H_{*}$  defines a new vector
field $\Gamma'\in \mathfrak{X}(T\R^{n})$  which is written as
\begin{equation}
\label{eq:91}
\Gamma'={\mathcal F}H_{*}(X_{H'})=f'(E_{H})\Gamma,
\end{equation}
which is Hamiltonian with respect to the symplectic form
$\omega_{H}$  (Equation \ref{eq:89}) and the function 
\begin{equation}
\label{eq:92}
E_{H'}=({\mathcal F}H^{-1})^{*}(H'),
\end{equation}
i.e., 
\begin{equation}
\label{eq:93}
i_{\Gamma'}\omega_{H}=dE_{H'}.
\end{equation}
In conclusion we have proved that a $f$-oscillator, when described in a Lagrangian framework, corresponds to a vector field $\Gamma'$ which is a conformal vector field with respect to a Harmonic oscillator vector field, the conformal factor being a function of the energy of the (undeformed) oscillator. We will see later on how the similarities with conformal vector fields can be used to obtain interesting properties of the deformations.

\section{Alternative tangent bundle structures from 
dynamics} 
\label{sec:altern-tang-bundle}

%We address now the main goal of the paper: given a SODE \marginpar{?}vector field
%$\Gamma'$ on some carrier space $M$,  is it possible to define a
%suitable tangent bundle structure on $M$ which makes $\Gamma'$ a SODE? 

\subsection{Defining the tangent bundle}
{}From Theorem \ref{theoremTM} we know that the tangent bundle structure
is encoded in a pair of tensors satisfying some compatibility
conditions. We can consider different tangent bundle structures in the same 
 carrier space $M$    by
choosing different pairs of tensor fields $(S, 
\Delta)$ in $M$ satisfying the properties above. It is obvious that by changing the tangent
bundle structure will change the nature of vector fields on $M$ as
SODE.  Our goal now  is to find a suitable pair $(\hat S, \hat
\Delta)$  on the manifold $M=T\R^{n}$  (the total manifold of the original
tangent bundle)  such that $\Gamma'$ satisfies that  
\begin{equation}
	\label{eq:99}
	\hat S(\Gamma')=\hat \Delta.
\end{equation}

In order to do that we are going to choose $n$ functions $\{ Q^{1},
\cdots, Q^{n}\}$ of $M=T\R^{n}$  which will play the role of the new
coordinates, satisfying
\begin{equation}
	\label{eq:100}
	(dQ^{1}\wedge \cdots \wedge dQ^{n})(p)\neq 0,
\end{equation}
for each point of $B$ or, at least, on a dense submanifold $Q$ of
$M$. These functions identify 
the base manifold of our new tangent bundle. Notice that the level set
of these new coordinates must define a vector space, which will
correspond to the tangent space at the point defined by the values of
the coordinates. Now, having in mind the requirement of $\Gamma'$ to
be a SODE, we  define: 
\begin{equation}
	\label{eq:102}
	{\mathcal V}^{k}={\mathcal L}_{\Gamma'}Q^{k}, \qquad
	k=1, \cdots, n,
\end{equation}
where the set of functions $\{ {\mathcal V}^{1}, \cdots, {\mathcal
  V}^{n}\}$  play the role of the new velocities.   
This, of course, implies that the coordinates must have the property that
$\Gamma'$ has no fixed points on the submanifold, i.e. 
$
\Gamma'(p)\neq 0$, $ \forall p\in Q$.
If all the velocity functions are
functionally independent, the bundle structure is well defined and
$\Gamma'$ becomes a SODE with respect to the 
new structure. Thus we have to require that
\begin{equation}
	\label{eq:50}
	dQ^{1}\wedge \cdots \wedge dQ^{n}\wedge d{\mathcal V}^{1}\wedge \cdots
	\wedge d{\mathcal V}^{n}\neq 0 \qquad \forall p\in M.
\end{equation}

In conclusion, the vector field $\Gamma'$ will take the form:
\begin{equation}
	\Gamma'=\sum_k\left (\mathcal{V}^k\frac{\partial}{\partial Q^k}+\mathcal{F}^k \frac{\partial}{\partial \mathcal{V}^k}\right ),
\end{equation}
where 
\begin{equation}
	\mathcal{F}^k= \mathcal{L}_{\Gamma'}\mathcal{V}_k
	\end{equation}
This construction would hold true chart by chart with appropriate
transition functions. In our case,  having chosen a linear space the problem becomes much simpler, since there is no need of studying several charts.

\subsection{Simplest examples: Harmonic oscillators}
\label{sec:simpl-exampl-harm}
Consider the simplest case of a manifold $M=\R^4$, parametrized by
coordinates $\{ x_1, x_2, x_3, x_4\}$ and the vector field 
\begin{equation}
\label{eq:12}
	\Gamma=x_2\frac{\partial}{\partial x_1}-x_1\frac{\partial}{\partial x_2}+
	x_4\frac{\partial}{\partial x_3}-x_3\frac{\partial}{\partial x_4}.
\end{equation}
Quite obviously, if we consider as base submanifold the vector space
$Q=\left \{ (x_1, x_2, x_{3}, x_{4})\in M| x_{2}=0=x_{4} \right \}$ and, as velocity coordinates 
$$
\mathcal{V}_1=\mathcal{L}_\Gamma x_1=x_2, \qquad \mathcal{V}_2=\mathcal{L}_\Gamma x_3=x_4,
$$
the vector $\Gamma$ becomes a SODE:
\begin{equation}
\Gamma=\mathcal{V}_1\frac{\partial}{\partial Q_1}+
\mathcal{V}_2\frac{\partial}{\partial Q_2}-Q_1\frac{\partial}{\partial
  \mathcal{V}_1} -Q_2\frac{\partial}{\partial \mathcal{V}_2}. 
\end{equation}

Other admissible choices, such as
$$
Q_B=\left \{ (x_1, x_2, x_{3}, x_{4})\in M| x_{2}=0=x_{3} \right \},
$$
$$
Q_C=\left \{ (x_1, x_2, x_{3}, x_{4})\in M| x_{1}=0=x_{4} \right \},
$$
$$
Q_D= \left \{ (x_1, x_2, x_{3}, x_{4})\in M| x_{1}=0=x_{3} \right \}
$$
would have defined analogous results, with choices for the velocities as
$\mathcal{V}_1=x_2$ and $\mathcal{V}_2=-x_3$ for $Q_B$, 
$\mathcal{V}_1=-x_1$ and $\mathcal{V}_2=x_4$ for $Q_C$ and
$\mathcal{V}_1=-x_1$ and $\mathcal{V}_2=-x_3$ for $Q_D$.

Had we considered instead a vector field of the form
\begin{equation}
	X=x_2\frac{\partial}{\partial x_1}-x_1\frac{\partial}{\partial x_2},
\end{equation}
it would be impossible to determine a suitable tangent bundle
structure making  it to be a SODE vector field . Indeed, it is immediate that it is
not possible to determine a two dimensional vector space $Q$ with the
required properties: 
\begin{itemize}
		\item If $Q$ cannot contain fixed points of the
                  vector field, the only possible choice is the linear
                  span of $x_1$ and $x_2$. 
		\item In that case, the velocity functions obtained as
		$\mathcal{V}_1=\mathcal{L}_XQ_1=x_2$ and
                $\mathcal{V}_2=\mathcal{L}_XQ_2=-x_1$ are not functionally
                independent of the coordinates of $Q$. 
	\end{itemize}

\subsection{An interesting example: conformal deformations 
of SODE   vector fields} 
\label{sec:an-inter-exampl-1}
Let us consider now as a particular example the case of a vector field
which is obtained from a SODE on a certain tangent bundle by a
conformal factor $f$, i.e., a vector field of the form 
\begin{equation}
	\Gamma'=f\,\Gamma= f (q,v)\left ( v^k\frac{\partial}{\partial q^k}+F^k(q,v)\frac{\partial}{\partial v^k} \right ),
\end{equation}
where $(q^k, v^k)$ is a set of coordinates adapted to a certain
tangent bundle structure defined on the carrier space $M$.  

In this situation, we want to identify a new tangent bundle structure
on $M$ which makes the conformal vector field $f\,\Gamma$ to be a SODE.

 Let us consider first the simpler case where $f$ is a constant of the
motion defined by $\Gamma$. If $\Gamma$ is not singular on $Q$, a
simple choice would be:  
\begin{equation}
\label{eq:39b}
Q^{k}=q^{k}, \qquad {\mathcal V}^{k}=\mathcal{L}_{\Gamma'} Q^k= f v^{k},
\end{equation}
where $(q^k, v^k)$ is the set of coordinates associated to the
trivialization of the first bundle structure. By construction the new 
coordinates are functionally independent and hence they satisfy
Equation \eqref{eq:50}. 
Tensors written 
as
\begin{equation}
S=dQ^\alpha\otimes \frac{\partial}{\partial {\mathcal V}^\alpha}, \qquad 
\Delta= {\mathcal V}^\alpha\frac{\partial}{\partial {\mathcal V}^\alpha},
\end{equation}
satisfy  Theorem \ref{theoremTM} and therefore they define a new tangent bundle structure on the manifold.

Furthermore, as $f$ is a constant of the motion for $\Gamma$, it is
immediate that the expression of $\Gamma'$ in the new coordinates
reads: 
\begin{equation}
\Gamma'={\mathcal V}^{k}\frac{\partial}{\partial {\mathcal Q}^k}+ fF^k\frac{\partial}{\partial {\mathcal V}^k},
\end{equation}
and therefore it is a SODE for the new bundle structure.

In the more general case where  $f$ is not a constant of motion for   $\Gamma$, the
situation is similar but  the expression of $\Gamma$ in the new variables becomes more
complicated: 
\begin{equation}
\label{eq:generalSODE}
	\Gamma'={\mathcal V}^{k}\frac{\partial}{\partial {\mathcal
            Q}^k}+ \tilde F^k\frac{\partial}{\partial {\mathcal V}^k}, 
\end{equation}
where 
\begin{equation}
\label{eq:11}
	\tilde  F^k= \mathcal{L}_{\Gamma'}\mathcal{V}^k=\Gamma'(f)v^k+fF^k= f
        v^kv^j \frac{\partial f}{\partial  q^j}+fv^kF^j\frac{\partial
          f}{\partial v^j}+f^2F^k. 
\end{equation}

\begin{example}
\label{sec:an-inter-exampl}
  Let us consider the simple case of the vector field  
  \eqref{eq:12} and a conformal function of the form
  \begin{equation}
    \label{eq:13}
    f=f\left ( (x_{1}^{2}+x_{3}^{2})(x_{2}^{2}+x_{4}^{2}) \right
    )=f\left (
      (Q_{1}^{2}+Q_{2}^{2})(\mathcal{V}_{2}^{2}+\mathcal{V}_{4}^{2})
    \right  ), 
  \end{equation}
together with  the first choice for the tangent bundle structure in
Section \ref{sec:simpl-exampl-harm}. Such a function depends thus on
the square of the angular momentum of the two-dimensional system. If
we redefine the tangent bundle structure to make it a SODE, the
resulting vector field will read
\begin{equation}
  \label{eq:14}
  \Gamma'=f\,\Gamma={\mathcal V}^{k}\frac{\partial}{\partial {\mathcal
      Q}^k}-f Q^{k}\frac{\partial}{\partial {\mathcal V}^k}.
\end{equation}
On the level sets of the angular momentum, this vector field
corresponds to a harmonic oscillator but where now the frequency
depends on the angular momentum.  We will see below that the situation
is similar to what happens with a $f$--oscillator, but the integral
curves change their frequencies with respect to different level sets.
\end{example}

Let us consider now two relevant applications of this construction to
physically interesting problems. In next Section, we will study the
relation between them. 

\subsection{Application 1: the closed orbits of the Kepler problem}

\label{sec:kepler-problem}

Let us consider now the classical Kepler problem in three
dimensions and let us consider the well known procedure used to regularize the corresponding vector field by defining
a lift to a four-dimensional configuration space by means of the
Kustaanheimo-Stiefel map (see
\cite{DAvanzo2004a,kustaanheimo1965perturbation}) for the details. 

\subsubsection{Kustaanheimo-Stiefel map}

The  $\KS$ map is explicitly defined as the  map
$\KS:\mathbb{R}_{0}^{4}\to \mathbb{R}_{0}^{3}$ 
\begin{equation}
	\label{eq:21}
	\KS(y^{0},y^{1},y^{2},y^{3})=(x^{1},x^{2},x^{3})
\end{equation}
given by 
\begin{equation}
	\label{eq:22}
	\begin{array}{rcl}
	x^1&= &2(y^0\,y^1+y^2\,y^3)\cr
x^2&= &2(y^0\,y^2-y^1\,y^3)\cr
x^3&= &(y^0)^2+(y^3)^2-(y^1)^2-(y^2)^2.
\end{array}
\end{equation}

This transformation can be seen  as an  extension to
$\mathbb{R}_{0}^{4}$ and $\mathbb{R}_{0}^{3}$ of the Hopf fibration
$\pi:S^{3}\to S^{2}$.  Indeed, we can identify $\mathbb{R}_{0}^{4}$
with $\R_{+}\times S^{3}$ and $\R_{0}^{3}$ with $\R_{+}\times S^{2}$.
Consider for instance the representation of
points in $\mathbb{R}_{0}^{4}\eqsim \mathbb{R}_{+}\times S^{3}$ as:
\begin{equation}
	\label{eq:26}
	 \mathbf{g} \mapsto \mathbf{g}\, \sigma_{3}\,\mathbf{g} ^{\dagger}=-\vec x \cdot \vec \sigma,\qquad  \mathbf{g}=\left (\begin{array}{cc}
		y^{0}+i\,y^{3} & y^{2}+i\,y^{1}\\
		-y^{2}+i\,y^{1} & y^{0}-i\,y^{3}
	\end{array}
	\right ) =R\,\mathbf{s},
\end{equation}
 where $\vec\sigma=(\sigma_1,\sigma_2,\sigma_3)$
 is the set of the three Pauli matrices, $R=
\sqrt{\sum_{\mu}(y^{\mu})^{2}}\in \mathbb{R}_{+}$ is the
distance to the origin and $\mathbf{s}\in S^{3}$ represents a 
parametrization of the three dimensional sphere of radius one by means
of determinant one unitary matrices as
\begin{equation}
\mathbf{s} =
\begin{pmatrix}
\alpha&\beta\\ - \beta^*&\alpha^*
\end{pmatrix}, \qquad \det \mathbf{s}=\vert\alpha\vert ^2
 +\vert\beta\vert ^2 =1. \label{bs}
\end{equation}

Consider also a parametrization of the unit sphere $S^{2}$  in terms of
matrices in  the Lie algebra
$\mathfrak{su}(2)$,
\begin{equation}
	\label{eq:27}
	q\in S^{2}\to q= -\vec x  \cdot \vec \sigma=
	\left (
	\begin{array}{cc}
		-x^{3} & -x^{1}+i\,x^{2}\\
		-x^{1}-i\,x^{2}& x^{3}
	\end{array}
	\right ); \qquad (x^{1})^{2}+(x^{2})^{2}+(x^{3})^{2}=1.
\end{equation}

A classical construction as the Hopf map $\pi:S^{3}\to S^{2}$
(\cite{Hopf1931,Urbantke2003})  is  written by using the previous
parametrization as 
\begin{equation}
	\label{eq:25}
	\pi (\mathbf{s})=\mathbf{s}\,\sigma_{3}\,\mathbf{s}^{\dagger},
\end{equation}
where as $\det \pi (\mathbf{s})=-1$,  we see that $\pi (\mathbf{s})$
describes a point of the sphere $S^2$.  If we replace $\mathbf{s}$
with a matrix as in  (\ref{eq:26}) we extend this map to  a map
$\mathbb{R}_{0}^{4}\approx 
\mathbb{R}_{+}\times S^{3} \to \mathbb{R}_{0}^{3}\approx
\mathbb{R}_{+}\times S^{2}$ as follows:
\begin{equation}
	\label{eq:KS}
\KS : \mathbf{g} \mapsto \mathbf{g}\, \sigma_{3}\,\mathbf{g} ^{\dagger}=-\vec x \cdot \vec \sigma,\qquad  \mathbf{g}=\left (\begin{array}{cc}
		y^{0}+i\,y^{3} & y^{2}+i\,y^{1}\\
		-y^{2}+i\,y^{1} & y^{0}-i\,y^{3}
	\end{array}
	\right )
\end{equation}
where now $
(x^{1})^{2}+(x^{2})^{2}+(x^{3})^{2}=r^{2}\in \mathbb{R}_{+}$.
The corresponding expression for the coordinates is
Equation (\ref{eq:22}).  Notice that the transformation
$\mathbb{R}_{0}^{4}\to \mathbb{R}_{0}^{3}$ satisfies 
$$
R^{4}=\left
((y^{0})^{2}+(y^{1})^{2}+(y^{2})^{2}+ 
(y^{3})^{2}\right )^{2}=(x^{1})^{2}+(x^{2})^{2}+(x^{3})^{2}=r^{2}.
$$
Notice that the unitary matrices defined as $s_{\phi}=\exp(\phi
\sigma_{3})$ preserve $\sigma_{3}$ and defines the set of
transformations projecting on the identity.

The tangent map of the covering map $\pi$,
$$T\pi:T(SU(2,\mathbb{C})\times \mathbb{R}_+)\to T(S^2\times \mathbb{R}_+)\,,$$
 is found to be given by 
\begin{equation}
\begin{array}{rcl}
v^1&=&2(y^0\,u^1+u^0\, y^1+u^2\,y^3+y^2\,\, u^3)\\
v^2&=&2(y^0\,u^2+u^0\, y^2-u^3\,y^1+y^3\, u^1)\\
v^3&=&2(y^0\,u^0+ u^3\,y^3-y^1\,u^1-y^2 \,u^2).
\end{array}
\end{equation}

By using this fact
it is immediate to define a Lagrangian system on $T\mathbb{R}_{0}^{4}$
whose solutions project on the solutions of the Euler-Lagrange
equation of the Kepler problem (see \cite{DAvanzo2004a}). The
corresponding Lagrangian reads:
\begin{equation}
	\label{eq:23}
	{\mathcal L}=2R^{2} \left (\dot R^{2}+R^{2}(\dot \theta_{1}^{2}+\dot
	\theta_{2}^{2}+\dot \theta_{3}^{2}) \right )+\frac
	k{R^{2}}=2R^{2}((v^{0})^{2}+(v^{1})^{2}+(v^{2})^{2}+(v^{3})^{2})+\frac k{R^{2}},
\end{equation}
where $R^{2}=(y^{0})^{2}+(y^{1})^{2}+(y^{2})^{2}+(y^{3})^{2}$ and $\theta_{k}$
are the left-invariant 1-forms on $SU(2)$ corresponding to the
relation
\begin{equation}
	\label{eq:24}
	s^{-1}ds=i \sum_{k}\sigma_{k}\theta_{k}  , \quad s\in SU(2)
\end{equation}

The advantage of using the $\KS$ map lies on the simplicity to define
the submanifold $\Sigma_{0}\subset T\mathbb{R}_{0}^{4}$ which is to be
mapped on the desired reduced dynamics. In this case, the submanifold
corresponding to Kepler solutions is written as:
\begin{equation}
	\label{eq:29}
	\Sigma_{0}=\left \{\, \, p \in T\mathbb{R}_{0}^{4}\mid  \dot \theta_{3}(p)=0
	\,\, \right\}
\end{equation}

\subsubsection{Redefining the tangent bundle structure}
\label{sec:newbundle}
After the unfolding procedure of the previous
section, we obtained the Euler-Lagrange  vector field $\Gamma\in \mathfrak{X}(T\R_{0}^{4})$
for the Lagrangian function 
\begin{equation}
	\label{eq:52}
	L=2R^{2}v^{2}+\frac g{R^{2}},
\end{equation}
which reads
\begin{equation}
	\label{eq:35}
	\Gamma=v^{k}\frac{\partial}{\partial y^{k}}+\left (
	\frac{v^{2}y^{k}}{R^{2}}- \frac{gy^{k}}{2R^{6}}-\frac{2\vec y \vec
		v}{R^{2}}v^{k}\right )\frac{\partial}{\partial v^{k}}. 
\end{equation}
This vector field projects on the Kepler vector field on $T\R_{0}^{3}$ via
reduction (see \cite{DAvanzo2004a} for details).
If we use the expression of the energy of the Lagrangian system on
$\R^{4}$
\begin{equation}
	\label{eq:51}
	E(y,v)=2R^{2}v^{2}- \frac{g}{R^{2}}
\end{equation}
we  can identify the level sets of $E(y,v)$
\begin{equation}
	\label{eq:36}
	\Sigma(E)=\left \{(y^{k}, v^{k})\in T\R_{0}^{4} \,\,\left  |  \right .
	E(y,v)=E  \right  \},
\end{equation}
which will prove to be very useful.

Notice that this vector field  (\ref{eq:35}) is not complete, exactly
as it happens with the three dimensional problem. But we know from
Section \ref{sec:repar-vect-fields} that it is possible to define a 
new vector field $\Gamma'=f\Gamma$, conformally related to $\Gamma$ by a suitable function $f\in
C^{\infty}(TR_{0}^{4})$,
 such that it is complete. Despite this,  we  do not want to 
loose the property of projecting on the Kepler problem in three
dimensions. Therefore, we have to look for a vector field
\begin{equation}
	\label{eq:31}
	\hat \Gamma=f \Gamma,
\end{equation}
where $f$ is a suitably chosen non-vanishing function, in such a way
that $T\KS(\hat \Gamma)=T\KS(\Gamma)$ and $\hat \Gamma$ is complete. 
Again, under such conformal transformation, the vector field
$\hat \Gamma$ may lose its nature as a SODE, but  it is
possible to redefine the tangent bundle structure of $T\R^{4}_{0}$ in
such a way that it becomes a SODE again.
Indeed, in \cite{DAvanzo2004a} it was proved that  a  suitable 
function and the corresponding redefinition of the tangent bundle
structure  is given by
\begin{equation}
	\label{eq:39}
	f=2R^{2}, \qquad Q^{k}=y^{k}; \qquad {\mathcal V}^{k}=2R^{2}v^{k},
\end{equation}
In local coordinates, this change translates as:
\begin{equation}
	\label{eq:46}
	T\R_{0}^{4}\ni (y^{k}, v^{k})\mapsto (Q^{k}, {\mathcal V}^{k})=
	\left \{
	\begin{array}{ll}
		Q^{k}&= y^{k}\\
		{\mathcal V}^{k}&=2R^{2}v^{k}
	\end{array}
	\right .
\end{equation}
where $R^{2}=\sum_{k=0}^{3}(y^{k})^{2}=\sum_{k=0}^{3}(Q^{k})^{2}$.  

Furthermore, it is immediate to prove that, restricted to a suitable submanifold, the vector field $\hat \Gamma$ coincides with the vector field of the harmonic oscillator.  
Indeed, let us consider the coordinate expression of the vector field
$\hat \Gamma$ with respect to the new bundle structure:
\begin{equation}
	\label{eq:55}
	\hat \Gamma={\mathcal V}^{k}\frac{\partial}{\partial Q^{k}}+ 2\left (
	\frac{{\mathcal V}^{2}}{2R^{2}}-\frac g{R^{2}}\right )Q^{k}\frac{\partial}{\partial {\mathcal V}^{k}}.
\end{equation}

Notice that the factor 
$$
\left (
\frac{{\mathcal V}^{2}}{2R^{2}}-\frac g{R^{2}}\right )
$$
is equal to the  energy function (\ref{eq:51}) written in the new coordinates.
Therefore the submanifold $\Sigma(E)$ is written now as 
\begin{equation}
	\label{eq:56}
	\Sigma(E)=\left \{(Q^{k}, {\mathcal V}^{k})\in T\R_{0}^{4} \,\,\left  | \left (
	\frac{{\mathcal V}^{2}}{2R^{2}}-\frac g{R^{2}}\right )=E<0 \right . \right  \}.
\end{equation}
Notice that this submanifold can also be considered as the level set
of a harmonic oscillator, with frequency $\omega=2E$ and energy equal
to $g$:
\begin{equation}
	\label{eq:62}
	\Sigma(E)=\left \{(Q^{k}, {\mathcal V}^{k})\in T\R_{0}^{4} \,\,\left  | 
	\frac{{\mathcal V}^{2}}{2}+| E  | R^{2}= g \right . \right  \}.
\end{equation}

The restriction of $\hat \Gamma$ to each $\Sigma(E)$ reads now
\begin{equation}
	\label{eq:57}
	\hat \Gamma |_{\Sigma(E)}={\mathcal V}^{k}\frac{\partial}{\partial Q^{k}}-
	2|E|Q^{k}\frac{\partial}{\partial {\mathcal V}^{k}}.  
\end{equation}
This is the vector field corresponding to a harmonic oscillator of
frequency $\sqrt{2|E|)}$. It is important to stress that 
we need a different harmonic oscillator for each Kepler energy to
accommodate a frequency depending on the energy.
The set of all
closed orbits of the Kepler problem would require then  a family
of different harmonic oscillator models, one oscillator  for each value of the energy.
This result is in agreement with classical results as
\cite{Cushman1997,Gyorgyi1968,Ligon1976,Marle2012} obtained in the Hamiltonian
formalism.

\subsection{ Application 2: Redefining the tangent bundle for
  $f$--oscillators}
\label{sec:appl-2:-redef}
We saw above how we can encode the notion of a f-oscillator in the
Lagrangian formalism, simply by using the Legendre transform and the
symplectic formulation of Lagrangian mechanics. 
But it is important to notice that, even if still symplectic,
$\Gamma'$ is no longer a SODE, as $\Gamma$ is. Our goal now is to study whether it is
possible to remedy this situation by defining a new tangent bundle structure on
$T\R^{n}$ which makes $\Gamma'$ a SODE. Notice that, from a formal point of view, the situation is completely analogous to the previous case (the regularized Kepler problem), since we have a (certain kind) of conformal vector field.

 In our
case we consider then, in a similar way to the case of the conformal Kepler vector field:
\begin{equation}
\label{eq:101}
Q^{k}=q^{k}
\end{equation}
Obviously the level set of the new coordinates become trivially vector spaces. On them, we choose velocities as:
\begin{equation}
\label{eq:102b}
{\mathcal V}^{k}={\mathcal L}_{\Gamma'}Q^{k}=f'(E_{H}(q,v))v^{k}, \qquad
k=1, \cdots, n;
\end{equation}
 It is important to remark that
 $f(E_{H})$ and its derivatives are preserved by the vector field
 $\Gamma$ and, therefore, also by the vector field $\Gamma'$.
Conditions (\ref{eq:100}) and (\ref{eq:50}) are trivially satisfied by
this choice and therefore if we consider the tensors
\begin{equation}
\label{eq:103}
\hat S= dQ^{k}\otimes \frac{\partial}{\partial {\mathcal V}^{k}},
\end{equation}
and
\begin{equation}
\label{eq:104}
\hat \Delta={\mathcal V}^{k}\frac{\partial}{\partial {\mathcal V}^{k}},
\end{equation}
they define a new tangent bundle structure on $M\simeq T\R^{n}$ which, by construction,
ensures that the vector field $\Gamma'$ is now  a SODE. Indeed, it is
straightforward to compute that, in the new coordinates, 
\begin{equation}
\label{eq:105}
\Gamma'=\sum_{k}\left ({\mathcal V}^{k}\frac{\partial}{\partial
	Q^{k}}-(f')^{2}Q^{k}\frac{\partial}{\partial {\mathcal V}^{k}} \right ).
\end{equation}

This is just the SODE corresponding to an oscillator with frequency
$\Omega=f'$. Notice that, as $f'$ is a constant of the motion, this
frequency is conserved by $\Gamma'$, even if it changes from orbit to 
orbit on $T\R^{n}$. 

Notice also that the original bundle structure of $T\R^{n}$ is
now lost, since new ``velocity'' coordinates depend on both the old
position and velocities. This transformation re-scales coordinates and velocities in
a different way, and therefore the energy level sets, which were
spheres in the original structure, become now ellipsoids. This fact will
be important in the next section.

Therefore, we have found how to encode the notion of $f$--oscillator
at the Lagrangian level: given a tangent bundle $TM$ and a harmonic
oscillator defined on it,  an $f$--oscillator is a conformal
transformation of the dynamical vector field $\Gamma\mapsto
\Gamma'=f\Gamma$ that can be combined with a change in the tangent bundle
structure of $TM$  in such a way that $\Gamma'$ becomes a SODE
representing an oscillator with non-uniform (but constant in time)
frequency $\Omega=f'(E_{H})$.
%\color{red}
%Can we give $\Gamma'$ a lagrangian description with respect to the new
%bundle structure? In 1D is always possible (see \cite{Currie1966967} ),
%the effect is seen on the potential 
%\color{black}

\section{Combining applications: writing conformal vector 
fields as   $f$-oscillators} 
\label{sec:space}
We have seen how $f$--oscillators, when considered from the Lagrangian
perspective, can be seen as a particular example of a conformal vector
field, where the original vector field corresponds to the undeformed
oscillator. What we want to study now is whether it is possible to
exploit this relation to  relate a given vector field $\Gamma$  and a
deformed oscillator, i.e., when can we find a suitable conformal
factor $g$ to make $\Gamma'=g\Gamma$ identical to a deformed
oscillator? In case it is not possible to find an identification, can
we establish some type of relation between both dynamical systems?

\subsection{The trivial case}
We saw in the previous section how the redefinition of the tangent
bundle structure allows us to write all conformal vector fields as
SODE vector fields in the form of Equation \eqref{eq:generalSODE},
while the particular case of a deformed oscillator leads to a vector
field as Equation \eqref{eq:105}. In this situation we may wonder: in
which situations does a conformal vector field represent the
Lagrangian version of a deformed oscillator?  

From our analysis in the previous section it is evident that there is
a trivial case where the answer to this question is positive: if we
consider the conformal vector field of the SODE representing a
harmonic oscillator by a function which is a function of the
oscillator energy we shall find a SODE vector field of the form 
\begin{equation}
\label{eq:general}
\Gamma'=g\Gamma=\mathcal{V}^k\frac{\partial}{\partial Q^k}-g(Q,
\mathcal{V}) Q^k\frac{\partial}{\partial {\mathcal V}^k}. 
\end{equation}

This expression is formally identical to \eqref{eq:105} and in both
cases they define closed orbits with frequencies depending on the
position in $M$, and the functions $g$ and $f$  may be chosen to
define the same frequency. If the level sets of both functions
coincide, the dynamics can be made to match perfectly. 

\subsection{Different functions: Souriau's space of motions}
\subsubsection{Formal relation of the vector fields}
Let us consider now a situation similar to the case of Equation
\eqref{eq:general} but where the function $g$ is now a constant of the
motion different from the energy, a natural example being the angular
momentum, as we presented in Example \ref{sec:an-inter-exampl}. Again
we obtain two SODE vector fields which are formally 
identical, but where submanifolds with constant frequencies are
different. We can conclude thus that the integral curves of both
vector fields will be related, but only the curves, since the
identification is given by the vector fields. 
Notice that the trajectories  with the same frequency are contained in
different submanifolds: those of Equation \eqref{eq:general} on the
level set of function $g$  and those of Equation
\eqref{eq:105} on the level set of the Harmonic oscillator energy. It
may happen that both functions are preserved by the vector
fields (for instance, Example \ref{sec:an-inter-exampl} ), but in any
case the frequency corresponding to 
each integral curve would be different for different points of the
carrier space.  Hence, in principle, the similarity of the two vector
fields can not be used to define a mapping between the carrier spaces.
Instead,  what the formal similarity of the two vector field allows us
to define is a mapping between the two sets of integral curves (one in the
level set of $g$  and the other in the level set of $f$) which have
the same value of the oscillator frequency. 

Notice that this relation depends only on the final expression of the
SODE vector field, and therefore we may think in relating the set of
trajectories of a deformed oscillator and those of any vector field which,
after a suitable conformal transformation, becomes the vector field of
a harmonic oscillator with frequencies that may depend on the
point. One example of such situation is the regularized Kepler
problem, that we analyze below.

\subsubsection{Souriau's space of motions}
J. M. Souriau introduced the concept of \textit{space of motions} to
represent the space of solutions of a given dynamical system (see
\cite{souriau1970structure}). In more precise terms, consider a
dynamical system defined by a vector field $X$ on a phase space
$M$. Consider also the extension of $M$ by $\R$ (to include the time),
that in Souriau's terminology is called the \textit{space of
  evolutions}. If the vector field $X$ is complete, the space of
evolutions is regularly foliated by the one-dimensional submanifolds
corresponding to the integral curves of $X$.  Notice that this space of
evolutions makes that even fixed points of $X$ become one dimensional
submanifolds, and this makes simpler the definition of a regular
foliation by these submanifolds. The \textit{space of motions}
$\mathcal{M}$ is defined as the corresponding quotient manifold, i.e.,
each point corresponds to a trajectory with a fixed parametrization.
Notice that, while the quotient above is well defined when $X$ is
complete, the resulting $\mathcal{M}$ being diffeomorphic to the phase
space, it may not be even a differential manifold if $X$ is not
complete.  A detailed analysis of the construction of the space of
motions for the Kepler problem can be found in \cite{Woodhouse1997}. 

\begin{figure}
\centering
\includegraphics[width=0.8\linewidth]{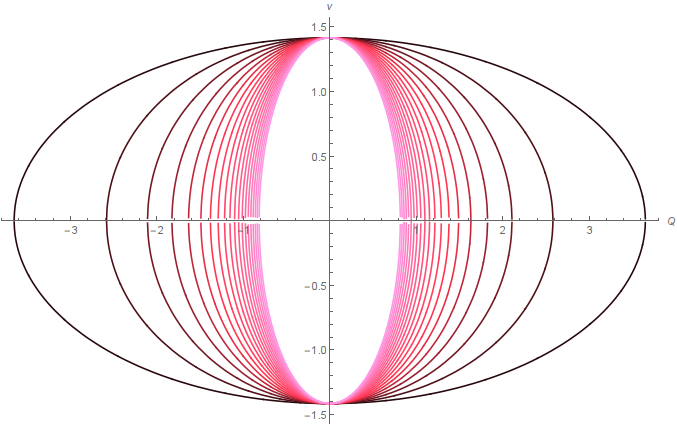}
\caption{Set of trajectories of the regularized Kepler problem
  (Equation \ref{eq:57}) with $g=1$ represented in the space $(Q,
  \mathcal{V})$ $Q$ being the modulus of the position and
  $\mathcal{V})$ the modulus of the velocity.} 
\label{fig:kepler-k}
\end{figure}

For instance, if we consider the set of closed orbits of the
regularized Kepler problem defined on $T\R^4$ by the vector field of
Equation (\ref{eq:57}) the space of motions corresponds to the set of
trajectories which can be represented in two dimensions by  Figure
\ref{fig:kepler-k}. Each of the curves represented by the lines in the
Figure (taking into account the parametrization) corresponds to one
point of $\mathcal{M}$.   

Notice also that a deformation of the type of a  $f$--oscillator, and
the resulting change in the tangent bundle structure of the system
produces also a transformation at the level of the space of
motions. Indeed, the change in the frequencies of the different orbits
is a clear indication of a transformation in the parametrization of
trajectories and hence in a mapping from one point in the space of
motions of the undeformed system to another point of the deformed one.
Our goal in the next section is to study whether a suitable
deformation function $f$ may achieve to define invertible mappings
between the spaces of motion of the deformed system and the space of
motions of the regularized Kepler problem.  

\subsection{Application: finding a relation for a $f$--oscillator
 and the  regularized Kepler problem}

In the previous sections we have been able to identify mechanisms
which allow us to construct alternative tangent bundle descriptions
for the regularization of the dynamical vector fields corresponding to
closed orbits of the Kepler problem and for the Lagrangian description
of the $f$--oscillators. Furthermore, in both cases the resulting
vector fields have the same formal expression and hence it is natural
to study whether it is possible to relate both descriptions. 

 If we consider the $f$--oscillator defined on
the same four dimensional configuration space as the regularized
Kepler problem of Equation (\ref{eq:57}), both vector fields will produce
trajectories of 4-dimensional harmonic oscillators with a given set of
frequencies depending on the points of the carrier space. The
submanifolds of points with the same frequencies will be different in
one case and the other, but sharing the same frequency is sufficient
to identify the corresponding \textit{motions} of the two vector fields.
 Therefore, we must  determine  a  function $f$ to produce a
$f$--oscillator such that the vector field (\ref{eq:105}) takes a set
of frequencies identical to those of the set orbits of the regularized
Kepler problem which project on closed orbits of the three dimensional
one.  
Thus the choice must satisfy
\begin{equation}
\label{eq:106}
\Omega=f'(|E|),
\end{equation}
Since the frequencies associated to each $\Sigma(E)$  
are equal to $\frac{\sqrt{2|E|^{3}}}g$ we have to choose a
deformation function $f$ satisfying
\begin{equation}
\label{eq:28}
f'(|E|)=\frac{\sqrt{2|E|^{3}}}{g} \Rightarrow f(|E|)=\frac 1{10g} (2|E|)^{5/2}+C.
\end{equation}
The constant can safely be fixed to zero. 

\begin{figure}
\centering
\includegraphics[width=0.8\linewidth]{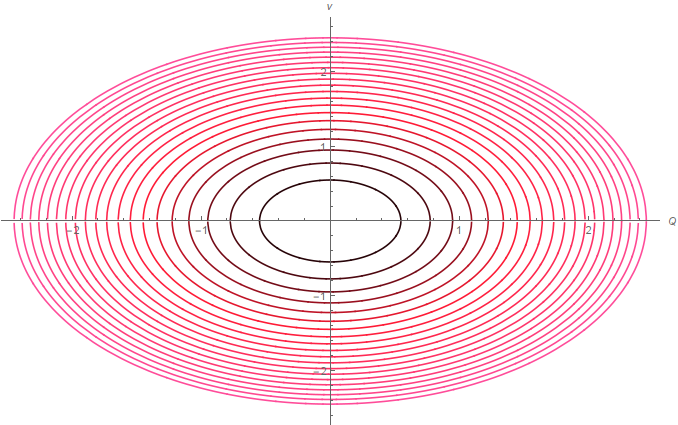}
\caption{Set of trajectories of the $f$--oscillator (Equation \ref{eq:105})
  with function equal to Equation (\ref{eq:28}) represented in the space
  $(Q, \mathcal{V})$ $Q$ being the modulus of the position and
  $\mathcal{V})$ the modulus of the velocity.} 
\label{fig:kepler-e}
\end{figure}

If we consider the set of motions of the corresponding $f$--oscillator we would get 
the set plotted in Figure \ref{fig:kepler-e}. Clearly there exists a
one-to-one correspondence between this space of motions and that of
Figure \ref{fig:kepler-k}, since the function $f$ has been chosen to
produce the same frequency for the periodic motion. 

With respect to such a system, the trajectory of the Kepler orbit with energy $E$
can be obtained by factorization of the $f$--oscillator orbit and the
correspondence between both orbit sets $\Psi$, i.e.

\begin{equation}
\label{eq:65}
\Phi_{t}^{K}=\Psi\circ \Phi_{t}^{f},
\end{equation}
where we represent by $\Phi_{t}^{K}$ and $\Phi_{t}^{f}$  the flows of
the corresponding systems at time $t$. As both vector fields are
complete, Equation (\ref{eq:65}) can be used to identify  the corresponding
spaces of motion in a trivial way.  

But it is also important to remark the importance of the changes in
the tangent bundle structures of both systems: without them it would
not have been possible to define the mapping above in a simple way. 

\section{Conclusions}
\label{sec:conclusions}

In this paper we have seen how the choice among alternative tangent
bundle structures for the carrier space $M$ of a certain dynamical vector
field $\Gamma$ can provide us with useful tools to describe interesting
properties of the corresponding dynamics. 
We have been able to identify general properties to identify the base
manifold by providing a set of coordinate functions $\{Q^{i}\}$ and
the corresponding submanifold $Q$ parametrized by them:
\begin{itemize}
\item $Q$ must have half the dimension of the carrier space
\item  the functions must be functionally independent
\item $Q$ must contain no fixed points of $\Gamma$ 
\item the level set of all the coordinate functions (i.e., a point
  $p\in Q$) must define a  submanifold of $M$ which is  a vector
  space (the corresponding   tangent space at the point of the base
  defined by $Q^{i}=p^{i}$) 
\item the set of velocity coordinates 
$$
\mathcal{V}^{k}= \mathcal{L}_{\Gamma}Q^{k},
$$
must be functionally independent among them and with respect to the
coordinate functions.
\end{itemize}
In these circumstances, we can define a tangent bundle structure that
makes $\Gamma$ a SODE.  In particular, we have seen
in Section \ref{sec:an-inter-exampl-1}
how common tools as the reparametrization of vector fields can be made
compatible with a redefinition of the bundle structure which
makes the reparametrized vector field a SODE. A particular example of
this construction is provided  in Section \ref{sec:kepler-problem} by
the reparametrized Kepler problem in four dimensions.  
A similar property has been proved for $f$--oscillators in Section
\ref{sec:appl-2:-redef}: the deformation by constants of the motion can
preserve the SODE property by a suitable redefinition of the tangent
bundle structure given by Equation (\ref{eq:101}) and Equation (\ref{eq:102b}). 
The usefulness of these alternative bundle structures  can be seen in
the last section: in some particular cases, it is possible to use the
resulting tangent bundle structures to define mappings between the
spaces of motions of two different systems, which may remain hidden
without them.  In particular we have been able to construct a
one-to-one mapping between the spaces of motion of the regularized
Kepler problem and of a suitable $f$-oscillator. Such a mapping
circumvents, partially, the obstruction defined by the energy-period theorem which
prevents the existence of a diffeomorphism between two systems having closed orbits
with different periods, as it happens with the Kepler problem and the
harmonic oscillator.

In the quantum domain it is possible to consider an analogue
construction by the redefinition of the hermitian product of the
Hilbert space. Indeed, in Quantum Mechanics it is possible to consider
alternative Hilbert space structures for the space of states and with
respect to them self-adjointness of operators may be adapted to
different situations (see, for instance, \cite{DAvanzo2005} for an
analysis  of hydrogen atom). In that case we may also search for
relations between different models with deformed oscillators, where
the deformation function may allow to match the equispaced spectrum of
the harmonic oscillator with the spectrum of the other model (for
instance the hydrogen atom). This is ongoing research that we expect
to publish soon.     

\section*{Acknowledgments}

The research of the first three authors has been financially supported
by the following Spanish grants: MICINN Grants FIS2013-46159-C3-2-P
and MTM2015-64166-C2-1-P, DGA Grant 24/1 and B100/13, and MECD Grant
FPU13/01587. 
  G.Marmo would like to acknowledge  the support provided by the Banco de
Santander-UCIIIM "Chairs of Excellence" Programme 2016-2017

%\section*{References}


\begin{thebibliography}{10}

\bibitem{josegeometry}
J.~F. Cari{\~{n}}ena, A.~Ibort, G.~Marmo, and G.~Morandi.
\newblock {\em {Geometry from Dynamics, Classical and Quantum}}.
\newblock Springer, 2014.

\bibitem{Carinena2007c}
J.F. Cari{\~{n}}ena, J.~Clemente-Gallardo, and Giuseppe Marmo.
\newblock {Reduction procedures in classical and quantum mechanics}.
\newblock {\em International Journal of Geometric Methods in Modern Physics},
  4(8):1363--1403, 2007.

\bibitem{crampin1983defining}
M~Crampin.
\newblock {Defining Euler-Lagrange fields in terms of almost tangent
  structures}.
\newblock {\em Physics Letters A}, 95(9):466--468, 1983.

\bibitem{Crampin1983}
M~Crampin.
\newblock {Tangent bundle geometry for Lagrangian dynamics}.
\newblock {\em Journal of Physics A: Mathematical and General}, 16:3755--3772,
  1983.

\bibitem{Crampin1985}
M.~Crampin and G.~Thompson.
\newblock {Affine bundles and integrable almost tangent structures}.
\newblock {\em Mathematical Proceedings of the Cambridge Philosophical
  Society}, 98(01):61--71, 1985.

\bibitem{Cushman1997}
RH~Cushman and JJ~Duistermaat.
\newblock {A characterization of the Ligon-Schaaf regularization map}.
\newblock {\em Comm. Pure Appl. Math.}, 50(8):773--787, 1997.

\bibitem{DAvanzo2004a}
A.~D'Avanzo and G.~Marmo.
\newblock {Reduction and unfolding: the Kepler problem}.
\newblock {\em Int. J. Geom. Meth. Mod. Phys.}, 2:83--109, 2005.

\bibitem{DAvanzo2005}
A.~D'Avanzo, G.~Marmo, and A.~Valentino.
\newblock {Reduction and Unfolding for Quantum Systems: the Hydrogen Atom}.
\newblock {\em Int. J. Geom. Meth. Mod. Phys.}, 2:1043--1062, 2005.

\bibitem{Filippo1989}
S.~De Filippo, G.~Landi, G.~Marmo, and G.~Vilasi.
\newblock {Tensor fields defining a tangent bundle structure}.
\newblock {\em Ann. de l'I.H.P., Section A}, 50(2):205--218, 1989.

\bibitem{gordon1969relation}
W.~B. Gordon.
\newblock {On the relation between period and energy in periodic dynamical
  systems}.
\newblock {\em J. Math. Mech}, 19:111--114, 1969.

\bibitem{Gordon1975}
W.~B. Gordon.
\newblock {Conservative dynamical systems involving strong forces}.
\newblock {\em Transactions of the American Mathematical Society},
  204:113--135, 1975.

\bibitem{Grifone1972a}
Joseph Grifone.
\newblock {Structure presque-tangente et connexions I}.
\newblock {\em Universit{\'{e}} de Grenoble. Annales de l'Institut Fourier},
  22(1):287--334, 1972.

\bibitem{Grifone1972b}
Joseph Grifone.
\newblock {Structure presque-tangente et connexions II}.
\newblock {\em Universit{\{}{\'{e}}{\}} de Grenoble. Annales de l'Institut
  Fourier}, 22(3):291--338, 1972.

\bibitem{Gyorgyi1968}
G.~Gy{\"{o}}rgyi.
\newblock {Kepler's Equation, Fock Variables, Bacry's Generators and Dirac
  Brackets.}
\newblock {\em Il Nuovo Cimento A Series 10}, 53(3):717--736, 1968.

\bibitem{Hopf1931}
H.~Hopf.
\newblock {{\"{U}}ber die Abbildungen der dreidimensionalen Sph{\"{a}}re auf
  die Kugelfl{\"{a}}che}.
\newblock {\em Math. Annalen}, 104:637--665, 1931.

\bibitem{kustaanheimo1965perturbation}
P~Kustaanheimo and E~Stiefel.
\newblock {Perturbation theory of Kepler motion based on spinor
  regularization}.
\newblock {\em Journal f{\"{u}}r Mathematik. Bd}, 218:204--219, 1965.

\bibitem{Lewis1955}
D.~C. Lewis.
\newblock {Families of Periodic Solutions of Systems having Relatively
  Invariant Line Integrals}.
\newblock {\em Proceedings of the American Mathematical Society},
  6(2):181--185, 1955.

\bibitem{Ligon1976}
T.~Ligon and M.~Schaaf.
\newblock {On the global symmetry of the classical Kepler problem}.
\newblock {\em Reports on Mathematical Physics}, 9(3):281--300, 1976.

\bibitem{Man'ko1997}
V.~I. Man'ko, G.~Marmo, E.~C.~G. Sudarshan, and F.~Zaccaria.
\newblock {f-Oscillators and Nonlinear Coherent States}.
\newblock {\em Physica Scripta}, 55(5):528--541, 1997.

\bibitem{Marle2012}
Ch.~M. Marle.
\newblock {A property of conformally Hamiltonian vector fields; Application to
  the Kepler problem}.
\newblock {\em The Journal of Geometric Mechanics}, 4(2):181--206, 2012.

\bibitem{Marmo1976}
G.~Marmo and A.~Simoni.
\newblock {Q-dynamical systems and constants of motion}.
\newblock {\em Lettere Al Nuovo Cimento Series 2}, 15(6):179--184, 1976.

\bibitem{Palais1957}
R.~Palais.
\newblock {A global formulation of the Lie theory of transformation groups}.
\newblock {\em Memoirs of the AMS}, 22, 1957.

\bibitem{souriau1970structure}
J~M Souriau.
\newblock {\em {Structure of Dynamical Systems}}, 
\newblock Birkhuser, Boston, 1997.

\bibitem{Urbantke2003}
H.K. Urbantke.
\newblock {The Hopf fibration—seven times in physics}.
\newblock {\em Journal of Geometry and Physics}, 46(2):125--150, 2003.

\bibitem{Woodhouse1997}
N.M.J. Woodhouse.
\newblock {\em {Geometric quantization}}.
\newblock Oxford University Press, 2nd edition, 1997.

\bibitem{Zhu1993a}
Chengjun Zhu and J.~R. Klauder.
\newblock {Classical symptoms of quantum illnesses}.
\newblock {\em American Journal of Physics}, 61(7):605, 1993.

\end{thebibliography}
\end{document}